% Uses lecproc.cmm

\input lecproc.cmm

\contribution{Dynamics, Structure, and Emission of Electron-Positron Jets}
\author{Amir Levinson}
\address{School of Physics and Astronomy, Tel Aviv 
University, Tel Aviv 69978, Israel}
                                                                                         
\abstract{
The theory of gamma-ray emission from e$^{\pm}$ jets and the implications for 
jet formation, dynamics and structure are reviewed.  In particular, possible carriers
of the jet's thrust on small scales, the transition from electromagnetic to particle
dominance in Poynting flux jets, formation of pair cascades, synchrotron emission 
by cascading pairs, and formation of shocks due to unsteadiness in the jet parameters
are considered, with emphasis on the  observational consequences.  Some recent progress
in modeling transient emission from blazars is also briefly discussed. }                     

\titlea{1}{Introduction}
\bigskip
There is little doubt that the gamma-ray emission seen from EGRET
blazars is highly anisotropic (see ,e.g., review by Schlickeiser 1996).  In most
models of high-energy emission from AGNs the energetic 
gamma-rays observed are attributed to emission processes in a relativistic jet pointing 
in our direction.  This view is strongly supported by the exclusive 
association of the EGRET AGN sources with compact radio sources (von Montigny
et al. 1995).  However, the physics of jet formation and dynamics
is not well understood.  Moreover, there are several unresolved issues related
to the emission from jets.  Further progress in our understanding of these systems
requires i) additional observations, particularly multi-waveband campaigns and coverage
of the 10-100 GeV band, and ii) theoretical tools which would enable interpretation of 
such data.  It is, therefore, important to study quantitatively various models 
of gamma-ray blazars.

This talk focuses on the physics of electron-positron jets.  Formation of e$^{\pm}$ jets, 
conceivable carriers of energy and momentum, and observational constraints on jet dynamics
and structure near the central engine are considered first.  Emission from leptonic jets
is considered next.  Finally, preliminary work concerning a specific mechanism 
for production of flares in blazars is discussed.                                                  
					
\titlea{2}{Observational characteristics of gamma-ray blazars}

The gamma-ray luminosities observed in blazars span a wide range, with the most 
powerful sources exhibiting isotropic 
luminosities during high states as high as 10$^{49}$ ergs s$^{-1}$.  The spectra in
the EGRET band can be well fitted  by power laws, with energy spectral indices in 
the range 0.7-1.5.  At least in two cases (Mrk 421 and Mrk 501) the spectrum extends into 
the TeV regime.  Some FSRQ exhibit spectral breaks (peaks) at a few MeV with a 
slope change $\Delta\alpha>1$ in some cases.  A second, low energy (radio to soft X-ray) 
spectral component peaking in the submm to IR regime is also characteristic to many blazars
(e.g., Brown et al. 1989).  This low energy component is commonly ascribed to
synchrotron cooling of relativistic electrons accelerated locally in the jet 
(Blandford \& K\"onigl 1979).  The high
energy component that peaks at MeV energies is probably produced by inverse Compton
emission of these  electrons.  However, the source of scattered photons is still 
unresolved, as discussed below.

Regarding the temporal behavior of blazars, doubling times as short as a few hours have 
been reported for some EGRET flares (e.g., Mattox et al. 1997), and an even shorter 
variability time scale has been inferred for Mrk 421 at TeV energies (Buckley et al. 
1996), implying very compact emission regions.  Recent observations reveal time lags 
of a few weeks to months between gamma-ray and radio outbursts 
(Reich et al. 1993; Zhang et al. 1994). 
In some sources there are also indications for correlations between the optical and 
gamma-ray emission (e.g., Maraschi et al. 1994; Wagner, 1996).  Such data can be used 
to impose constraints on the relative location of the emission regions at different 
energies, and perhaps on the emission mechanism. 

\titlea{3}{Electron-Positron jets}

\titleb{3.1}{The small scale structure}

The pair content of leptonic jets is limited at small radii by annihilation. 
For a conical jet with an opening angle $\sim \Gamma^{-1}$, where 
$\Gamma$ is the bulk Lorentz factor,
the maximum jet's thrust that can be carried by sub-to-
mildly relativistic pairs is given by (Blandford and Levinson 1995)
$$
L_{e}=\left({\sigma_T\over 2\pi\sigma_{ann}}\right)
\left({m_e\over m_p}\right)\left({r\over r_g}\right)\Gamma L_{Edd}
\simeq 5\times10^{45} \Gamma_{1} r_{16}\ \ \ {\rm ergs\  s^{-1}},
\eqno(1)
$$
where $r_g$ is the gravitational radius, $\Gamma_1=\Gamma/10$, and $r_{16}$ is the 
radial distance from the putative black hole in units of $10^{16}$ cm.  The enormous 
gamma-ray luminosities observed from the powerful
sources imply jet power of at least $10^{46}$ - $10^{47}$ ergs s$^{-1}$.
Consequently, if jet formation is completed close to the black hole (
at a distance of a few gravitational radii say) then 
the carrier of energy and momentum at radii below $r_{ann}\sim 10^{-2}$ pc
must be either baryons, Poynting flux, or ultra-relativistic pairs for which the 
annihilation cross section is sufficiently reduced by KN effects.
Alternatively, the jet may be accelerated and collimated over a range of radii
encompassing $r_{ann}$.  Hadronic jet models (Mannheim 1993; Dar \& Laor 1997) 
will not be considered here (see e.g., Celotti 1997; Mannheim 1997).  The 
possibility that the jet power is transferred outwards
by ultra-relativistic pairs requires most of the jet power to be dissipated below 
$r_{ann}$ in ERC models (which invoke the presence of external, dense radiation field)
if the jet accelerates to $\Gamma$ in access of that of the frame 
in which the radiation field is roughly isotropic.  Such a model can account for the 
MeV peak seen in several sources (since gamma-rays having energies below a few MeV can 
escape without being absorbed by pair production on the background photons, as 
discussed in \S 3.2 below).  However, the gamma-ray spectrum above the 
peak is anticipated, in this scenario, to be much steeper than those typically observed. 
Furthermore, this may be problematic for unified models in which the radio 
luminosities of extended radio sources are associated with the jet power on large scales.
The reason is that the observed luminosity of extended lobes is predicted to be much 
smaller than the anisotropic gamma-ray luminosities (i.e., after correcting
for beaming effects) inferred in blazars, in conflict with observations.

In the case of a cold e$^{\pm}$ beam the production rate of soft X-rays by the 
interaction of the cold electrons with the ambient radiation field is related
to the electron kinetic power, $L_e(r)$, through (Levinson 1996b), 
$$
{dL_X\over d\ln r}=L_e(r){r\over {l}_c},
\eqno(2)
$$
where 
$$
{l}_c/r\simeq 0.5 (\chi L_{x46})^{-1}r_{16}\Gamma^{-1}
\eqno(3)
$$
is the ratio of inverse Compton cooling length of streaming electrons
to jet radius.  Here $10^{46}L_{x46}$ ergs s$^{-1}$ is the luminosity of the background    
radiation, and $\chi$ the fraction of this luminosity that is intercepted by the 
jet.  For a reasonable choice of parameters we find that 
this ratio becomes smaller than unity below $r_{ann}$.  
Clearly, in order to avoid catastrophic radiative drag and hence X-ray overproduction (i.e.,
$L_X<L_j$) the fraction of energy flux carried by electrons (positrons) in 
the inner jet, which for an outflow consisting of purely e$^{\pm}$ plasma equals 
$\Gamma_A^{-1}$, $\Gamma_A$ being the Lorentz factor associated with the 
Alf\'ven speed of the outflow with respect to its rest frame,
must be smaller than the ratio of radiative cooling time to outflow time ${l}_c/r$.  
The constraint on $L_e$ might be even more stringent in sources in which the soft
X-ray luminosity is inferred to be much smaller than the jet power (e.g., Sikora 
et al. 1997).  The above conclusions may be substantially changed if the jet consists
of a relativistic core shielded by a slower, hot, Thomson thick outflow.

A scenario in which energy 
extracted from a spinning black hole is transfered outwards in the form of a 
Poynting flux jet which is collimated by a surrounding hydromagnetic wind emanating from 
an accretion disk, has been discussed recently (Blandford \& Levinson 1995).  In this model
the jet undergoes a transition from electromagnetic to particle dominance in the 
vicinity of the annihilation radius $r_{ann}$ (Levinson \& Blandford 1995).
The conversion of electromagnetic energy into pairs and X/gamma-rays can result 
from (Levinson 1996b) 
either the interaction of the cold $e^{\pm}$ beam with the ambient radiation,
or strong dissipation in the gamma-ray emitting region (beyond $r_{ann}$), e.g.,
due to the formation of dissipative fronts by unsteady jet injection (Romanova
\& Lovelace 1997; Levinson \& van Putten 1997).  In the former case, copious
pair production ensues once the jet is accelerated to bulk Lorentz factor in 
excess of $E_{thr}/m_ec^2$, the threshold energy above which the 
opacity to pair production on background photons exceeds unity.  For the
standard spectrum (Blandford and Levinson 1995) $E_{thr}\sim (m_ec^2)^2/E_{max}$
, where $E_{max}\sim 100$ KeV is the maximum cutoff energy of the scattered spectrum.   
The asymptotic Lorentz factor is then limited to $\Gamma\sim m_ec^2/E_{max}
\sim10$ (Levinson 1996b), compatible with that inferred from superluminal expansions
(Vermeulen \& Cohen 1994).  One problem with this mechanism is that it requires
the spectrum of the soft photons intercepted by the inner jet (but not by the 
gamma-ray emitting jet) to be sufficiently flat 
in order to avoid X-ray overproduction (Levinson 1996b).
In the latter case it is envisioned that continues fluctuations of the outflow steepen
into a train of shocks above the cooling radius. The shocks thereby created propagate 
along the jet and dissipate
a substantial fraction of the jet energy over the extended, gamma-ray emitting region.
The resultant spectrum then peaks in the MeV band, as explained below.  Frequent formation
of such shocks can lead to a slowly varying (quiescent) emission, whereas occasional
creation of a very intense front may lead to a gamma-ray flare, as discussed further
in \S 3.4. 

\titleb{3.2}{Synchrotron and inverse Compton emission}

As already mentioned above, the radio to UV/soft X-ray continuum spectrum is successfully
interpreted as synchrotron radiation by relativistic electrons (positrons)
accelerated in situ, while the high-energy spectral component is presumably due
to inverse Compton emission of these electrons.  The source of seed photons can be
either the synchrotron radiation itself (SSC mechanism,  e.g., K\"onigle 1981; Ghisellini 
\& Maraschi 1989; Bloom \& Marscher 1993), nuclear radiation that directly
enters or, alternatively, scattered (reprocessed) across the jet by surrounding gas 
(Dermer \& Schlickeiser 1993; Blandford \& Levinson 1995; Sikora, Begelman, \& Rees 1994; 
Marcowith, Henri, \& Pelletier 1995), or jet synchrotron emission reprocessed 
by the broad line clouds (Ghisellini \& Madau 1996).   
The ERC mechanism is likely to dominate in the powerful gamma-ray quasars if they
posses isotropic UV/X-ray luminosities as high as those typically observed in 
radio-quiet sources (e.g., Sikora, et al. 1997).  Moreover,
SSC models have difficulties explaining the high ratio of luminosities of the 
high-and low-energy spectral components often seen in the powerful blazars (Mannheim 1997; 
Sambruna et al., 1997; Sikora, et al. 1997).  This mechanism is more likely to be important in 
the weak BL Lac objects in which the luminosity of the underline nuclear radiation 
appears to be low. 
												     
The UV/soft X-ray background may also contribute a large opacity to pair 
production at small radii (Dermer \& Schlickeiser 1993; Sikora et al. 1994; 
Blandford \& Levinson 1995).  To be concrete, for a soft photon intensity typical to radio 
quiet quasars, the gamma-spheric radius below which the pair production opacity becomes 
larger than unity increases with gamma-ray energy and lies in the range 10$^{-3}$ to about 
0.1 pc at EGRET energies (Blandford \& Levinson 1995).  This imposes a constraint on the 
location of the gamma-ray emission region.     

In one-zone models (e.g., Sikora et al. 1994), the broad-band emission (with the possible 
exception of the radio emission) is assumed to originate from a small region where dissipation 
of the bulk energy predominantly takes place. (In the ERC version this region should be located 
far enough from the central source to avoid attenuation of the highest
energy gamma-rays observed.)  Correlations between the fluxes at different 
energies over a broad energy range may then be naively anticipated,  although situations
wherein variations in the energy distribution of emitting electrons may lead to a different 
behavior can be envisioned.  Such a prediction appears to be consistent 
with the claimed correlations between optical and gamma-ray emission (Wagner 1996). 
Unfortunately, this observation is not discriminatory since such correlations are predicted 
also by inhomogeneous pair cascade models, given the EGRET sensitivity 
(Levinson 1996a).  The reported delays between 
gamma-ray and radio flares (Reich et al. 1993) are not in conflict with the one-zone model  
provided that the emission region is located well within the radio core.  If the emission 
region is at a distance of $10^{17}$ - $10^{18}$ cm from the central source, as suggested by
Sikora et al. (1994), then the gamma-ray spectrum should exhibit a high-energy cutoff
in the range 10-100 GeV.  This energy band is presently uncovered by any 
instrument.  It is hoped that the next generation gamma-ray telescope (like GLAST) will 
help elucidating the spectrum of gamma-ray blazars in this range.
 
In the inhomogeneous pair cascade models (Blandford \& Levinson 1995;
Marcowith, et al. 1995), which assume continues dissipation and electron acceleration 
along the jet, the observed gamma-rays at a given energy are created near the 
corresponding gamma-spheric radius through pair
cascades.  As a result, the emitted gamma-ray spectrum is produced over a large range of 
jet radii, with higher energy gamma-rays coming from larger radii, and reflects 
essentially the intensity of the ambient radiation as well as the variation of electron 
injection rate with jet radius.  The energy
distribution of the radiating electrons is determined by the cascade process and is 
highly insensitive
to the injected electron spectrum, provided that electron acceleration is efficient.  
In contrast to the one-zone models, inhomogeneous pair cascade models predict 
spectral evolution during gamma-ray flares, with slower (or later) variations of the 
gamma-ray flux at higher energies.  The detection of such a spectral evolution in quasars
requires coverage of energy range broader than that covered by EGRET with a better 
sensitivity, and should be one of the objectives of future missions.
The simultaneous X-ray/TeV flare and the lack of 
significant changes in the EGRET flux seen in Mrk 421 (Macomb et al. 1995; Takahashi 
et al. 1996) is problematic for this model.  The absorption
by pair production on the background photons gives rise to a steepening of the gamma-ray 
spectrum at energies above about $(m_ec^2)^2/E_{max}\sim 10$ MeV ($E_{max}$ is the high 
energy cutoff of the ambient radiation mentioned above) and, therefore, can account 
quite naturally for the MeV bump.  
												      
The cascading pairs are also responsible
for the synchrotron spectrum.  A detailed analysis of synchrotron emission from 
inhomogeneous pair cascade jets (Levinson 1996a) shows that the radio to UV spectra
observed typically in blazars can be reproduced by the model quite naturally, provided
that the product of pair injection rate and magnetic field declines sufficiently 
steeply with radius (steeper than $r^{-3}$).  The turnover from flat to a 
steeper power law spectrum results, in the model, from the strong suppression of the
synchrotron emissivity below $r_{ann}$, owing to rapid pair annihilation (see \S 3.1). 
A second break at higher frequencies (observed in some sources) can be reproduced by 
controlling the maximum electron injection energy.  Further, for typical parameters
the radio (GHz) photospheres are located well beyond the EGRET gamma-spheres whereas the 
submm to optical emission region coincides with the jet section where EGRET gamma-rays are 
produced.  Given the sensitivity of EGRET, the latter is consistent with the 
optical/gamma-ray correlations discussed above.

\titleb{3.3}{Local electron acceleration and maximum injection energy}

The maximum energy attainable by an electron depends on the acceleration rate.  
Shock acceleration can give rise to a maximum acceleration rate on the order 
of the gyro-frequency of accelerated particle (Blandford and Eichler 1987; 
Kirk 1997).  For electrons (positrons) this yields a maximum Lorentz factor, as 
measured in the comoving frame,
$$
\gamma_{max}\simeq 10^8(\eta/B)^{1/2}(1+{\cal E})^{-1/2},
\eqno(4)
$$
where $\eta$ is the acceleration rate in units of the electron gyro-frequency, and
$$  
{\cal E}={U_{x}\over U_B}\simeq 6\times10^5 {\chi L_{x46}\Gamma_1^2\over r_{16}^2B^2}
\eqno(5)
$$
is the ratio of comoving energy densities of scattered radiation and magnetic field.
Note that this ratio is independent of radius if $B\propto r^{-1}$.
For a reasonable choice of parameters (cf. Levinson 1996a) we find that the 
maximum electron energy is not likely to exceed a few TeV or so 
in the powerful sources.  Higher energies may 
be attainable in faint BL Lac objects provided that the magnetic
field is sufficiently weak.  The fact that Mrk 421 and Mrk 501 have been detected
at TeV energies implies that at least in these sources electron injection must be 
very effective. 
It is not known whether the high energy spectrum of FSRQ extends
into the TeV regime.  TeV detections of FSRQ would impose severe constraints on ERC
models.  Observations of FSRQ in the energy range 10 GeV to a few hundred GeV, where 
absorption by the intergalactic IR background is strongly suppressed, can provide
valuable information regarding the in situ acceleration mechanism and the location of the
gamma-ray emission region. 

\titleb{3.4}{Time dependent models}

Various episodes may lead to time variability of blazar emission.  For example,
sudden changes in particle injection rate and/or magnetic field, changes in the
bulk speed, or temporal changes of the intensity of ambient radiation in ERC
models.  Presumably, different mechanisms would give rise to different 
characteristics of the time dependent emission in blazars.  It is, therefore, 
desirable to explore different variability models, and compare model
predictions with observations.  Below, we briefly discuss a specific model of
transient jets.									   	  

Romanova \& Lovelace (1997) proposed a model of gamma-ray and VLA
flares in which fluid collision in a pointing flux jet leads to the formation of
radiating fronts propagating down the jet.  Under the assumption of rapid magnetic
field dissipation (and therefore slow expansion of the front) they derived
a set of differential equations governing the acceleration, heating and cooling of the 
front.  The solution of the system yields the predicted light curves.
In their treatment they ignored gamma-ray absorption on external, hard X-ray photons,
which is expected to be important at small radii as explained above,
and, therefore, obtained almost simultaneous flaring at frequencies above the synchrotron
self-absorption frequency.  Further, the energy distribution of shocked particles is assumed
a priori in their model.

The formation, evolution and structure of such fronts have been carefully examined recently
by Levinson \& van Putten (1997), using analytic and numerical approach.  By treating the 
magnetic field in the front as a free parameter they determined the dependence of the 
front parameters on the rate of magnetic field dissipation.  The distance from the injection
point at which disturbances steepen into shocks is found
to be roughly $c\Delta t\Gamma^2\Gamma_A^2/3$, where $\Delta t$ is the characteristic time 
change of the outflow parameters (of order the dynamical time in the injection zone).  This 
model can be extended (Levinson, in preparation), within the framework of the inhomogeneous 
pair cascade model, 
to incorporate radiative cooling and pair cascades self-consistently, by coupling the front 
equations with the equations governing the evolution of the pairs, gamma-rays and synchrotron
intensities in the front.  This would enable self-consistent calculations of light curves as 
well as spectral evolution during flares over a range encompassing radio to gamma-ray energies
under different conditions (e.g., magnetic field dissipation rate).

\titlea{}{Acknowledgment} 
I thank R.D. Blandford, M. Van Putten, R.V.E. Lovelace, and M. Romanova for useful discussions.
Support by Alon Fellowship is greatly acknowledged.      
											            
\vfill \eject

\centerline{\bf REFERENCES}
\bigskip

\noindent{Blandford R.D. \& K\"onigl, A. 1979, ApJ, 232, 34}

\noindent{Blandford R.D. \& Eichler, D. 1987, Phys. Rep., 154, 1}

\noindent{Blandford R.D. \& Levinson, A. 1995, ApJ, 441, 79}

\noindent{Bloom, S.D., \& Marscher, A.P. 1993, in Compton Gamma-Ray 
Observatory, eds. N. Gehrels \& M. Friedlander (New York: AIP), 578}

\noindent{Brown, L.M.J., et al. 1989, ApJ, 340, 129}

\noindent{Buckley, HH., et al. 1996, ApJ, 472, L9}

\noindent{Celotti, A. 1997, These proceedings}

\noindent{Dar, A., \& Laor, A. 1997, ApJ, 478, L5}

\noindent{Dermer, C.D., \& Schlickeiser, R. 1993, ApJ, 416, 458}

\noindent{Ghisellini, G., \& Maraschi, L. 1989, ApJ, 340, 181}

\noindent{Ghisellini, G., \& Madau, P. 1996, MNRAS, 280, 67}

\noindent{ Kirk, J. 1997, These proceedings} 

\noindent{ K\"onigle, A. 1981, ApJ, 243, 700}

\noindent{Levinson, A. \& Blandford R.D. 1995, ApJ, 449, 86}

\noindent{Levinson, A. 1996a, ApJ, 459, 520}

\noindent{Levinson, A. 1996b, ApJ, 467, 546}

\noindent{Levinson, A. \& Van Putten, M. 1997, ApJ, in press}

\noindent{Macomb, D.J., et al. 1995, ApJL, 449, L99}

\noindent{Manneheim, K. 1993, A\&A, 269, 67}

\noindent{Manneheim, K. 1997, preprint (astro-ph/9703184)}

\noindent{Maraschi, L., et al. 1994, ApJ, 435, L91}

\noindent{Marcowith, A., Henri, G., \& Pelletier, G. 1995, MNRAS, 277, 681}

\noindent{Mattox, J., et al. 1997, ApJ, 476, 692} 

\noindent{Reich, W. et al. 1993, A\&A, 273, 65}

\noindent{Romanova, M.M., \& Lovelace, R.V.E. 1997, ApJ, 475, 97}

\noindent{Sambruna, R.M., et al. 1997, ApJ, 474, 639}  
  
\noindent{Schlickeiser, R. 1996, SSRv, 75, 299}

\noindent{Sikora, M., et al. 1997, preprint}

\noindent{Sikora, M., Begelman, M., \& Rees, M.J. 1994, ApJ, 421, 153}

\noindent{Takahashi, T. et al. 1996, ApJL, 470, L89}

\noindent{Vermeulen, R.C., \& Cohen, M.H. 1994, ApJ, 430, 467}
 
\noindent{Wagner, S.J. 1996, ApJS, 120, 495}

\noindent{Zhang, Y.F., et al. 1994, ApJ, 432, 91}

\bye